\documentclass[10pt]{IEEEtran}
\pdfoutput=1

\usepackage{cite}
\usepackage{amsmath}
\usepackage{amsthm}
\usepackage{amssymb}
\usepackage{array}
\usepackage{algorithm}
\usepackage{algpseudocode}
\usepackage{mathtools}
\usepackage{graphicx}
\DeclareGraphicsExtensions{.pdf}

\theoremstyle{plain} \newtheorem{theorem}{Theorem}
\theoremstyle{plain} \newtheorem{proposition}{Proposition}
\theoremstyle{plain} \newtheorem{lemma}{Lemma}
\theoremstyle{plain} 			
\theoremstyle{remark} \newtheorem{remark}{Remark}
\theoremstyle{definition} \newtheorem{definition}{Definition}			

\begin{document}
\title{On Generalized Optimal Hard Decision Fusion}

\author{Mohammad Fayazur Rahaman, \IEEEmembership{Student Member, ̃IEEE,} and Mohammed Zafar Ali Khan, \IEEEmembership{Senior Member, ̃IEEE}
	\thanks{M.Fayaz and Prof. Zafar are with Indian Institute of Technology Hyderabad, India}
	\thanks{This research work was partly sponsored by Visvesvarya YFRF, Miety.}
}
\maketitle

\begin{abstract}
	In this letter we formulate a generalized decision fusion problem (GDFP) for sensing with centralized hard decision fusion.
	We show that various new and existing decision fusion rules are special cases of the proposed GDFP.
	We then relate our problem to the classical $0-1$ Knapsack problem (KP). Consequently, we apply dynamic programming to solve the exponentially complex GDFP in polynomial time. Numerical results are presented to verify the effectiveness of the proposed solution.
\end{abstract}


\section{Introduction}
A binary hypothesis sensing involves detecting the presense (hypothesis $H_1$) and absense ($H_0$) of the phenomenon being observed. 
Distributed sensing improves the reliability of the sensing decisions about the phenomenon. 
The local sensors compute their binary hard decisions independently ($u_i$) and forward the same on bandwidth constrained reporting channels to the fusion center (FC). The probability of detection ($P_D$, correctly declaring $H_1$) and probability of false alarm ($P_F$, incorrectly declaring $H_1$) at the FC are commonly used performance measures of the system.

Optimization of the distributed sensing with given sensor performance characteristics for Bayesian and Neyman-Pearson (NP) criterion has been studied in \cite{Chair1986,Hoballah1986,Thomopoulos1987,Thomopoulos1989,Hoballah1989,Varshney1997,Viswanathan1997}.
It is shown in \cite{Varshney1997,Viswanathan1997} that the problem of optimal hard decision fusion in a general NP setting is exponentially complex. In this letter we propose a polynomial time solution for a more generalized problem.

Chair-Varshney (CV) \cite{Chair1986} have derived an optimal, linear  fusion rule at the FC for the Bayesian test. With similar assumptions, randomized decision fusion rule is derived in \cite{Hoballah1986} using randomized LRT \cite{Trees2001} for the NP criterion. In \cite{Thomopoulos1987}, it is shown that the CV linear fusion equation simplifies to $K$-out-of-$N$ rule when the sensors are homogenous.
A simple $K$-out-of-$N$ voting rule is used in \cite{Zhang2009} and closed form expressions are derived for optimum $K$, $N$ and FC threshold for a special case of the Bayesian cost function. In a similar setting, optimal results are derived for erroneous reporting channel in \cite{Banavathu2015}. Performance comparison of $K$-out-of-$N$ and soft decision fusion rule over erroneous reporting channel is presented in \cite{Chaudhari2012}.
A person-by-person (PBPO) iterative approach to jointly optimize the decisions at the sensors and the fusion rule at FC is proposed in \cite{Hoballah1989} for the Bayesian criterion. A similar PBPO iterative approach is proposed in \cite{Chen2005,Chen2009} for the Bayesian and NP criterion for erroneous reporting channels. Particle swarm optimization	algorithm is used in \cite{Veeramachaneni2007} to jointly optimize the LRT thresholds at the sensors and the FC for the Bayesian cost function. In \cite{Peh2010} different scenarios for computing the thresholds, jointly and separately, are discussed with the objective to optimize the throughput of the system.
FC applies a linear weighted sum fusion rule on a multi-bit test statistics received from the sensors in \cite{Taricco2011,Quan2010} for NP criterion.

The outline of our letter is as follows:
In Section \ref{sec:sys_model} we explain the system model and the general fusion rule. The GDFP is formulated in Section \ref{sec:prob_form} and special cases are derived. In Section \ref{sec:sol_gen} we present dynamic programming based algorithm to solve the GDFP and provide simplified solutions to the special cases. Section \ref{sec:num_res} contains the numerical results, followed by conclusions in Section \ref{sec:con}.

\section{System Model}
	\label{sec:sys_model}
	\subsection{Heterogenous Sensors}
	We consider a system of $N$ sensors where each sensor is characterized by its average probability of detection $p_{d_i}$ and false alarm $p_{f_i}$ as:
	\begin{IEEEeqnarray}{lClCl}
		\label{eq:pfi}
		p_{d_i}&=&Pr\{u_i=1 \mid H_1\},\quad p_{f_i}&=&Pr\{u_i=1 \mid H_0\},
	\end{IEEEeqnarray}
	where $u_i \in \{1,0\}$ is the binary decision of the $i^{th}$ sensor indicating hypothesis $H_1$ and $H_0$ respectively. Following \cite{Chair1986}, we assume \{$p_{d_i},p_{f_i}\}, \: \forall i$ to be known.
	
	Define probability vectors $\mathbf{p_d} \triangleq\ [p_{d_{N-1}} \cdots p_{d_0}]$, $\mathbf{p_f} \triangleq\ [p_{f_{N-1}} \cdots p_{f_0}]$ and decision vector $\mathbf{u} \triangleq {[u_{N-1} \: \cdots \:  u_0]}$. For each sensing cycle, the FC receives $\mathbf{u}$ and generates a fused binary decision $u_{fc} \in \{1,0\}$.
				
	Assuming that the $u_i's$ are conditionally independent, the probability of occurrence of a specific decision vector at the FC under $H_0$ and $H_1$ is \cite{Varshney1997}:
	\begin{IEEEeqnarray}{lCl}
		\label{eq:Pdfm}
		Pr\{\mathbf{u} \mid H_1\} &=& \prod_{i=0}^{N-1} (p_{d_i})^{u_i} \cdot (\bar{p}_{d_i})^{1- u_i} \quad , \nonumber \\
		Pr\{\mathbf{u} \mid H_0\}
		 &=& \prod_{i=0}^{N-1} (p_{f_i})^{u_i} \cdot (\bar{p}_{f_i})^{1- u_i} \quad ,
	\end{IEEEeqnarray}
	where $\bar{p}_{d_i} = 1-p_{d_i}$ and $\bar{p}_{f_i} = 1-p_{f_i}$.
	
	A total of $M=2^N$ distinct decision vectors $\mathbf{u}_m$ are possible with index $m \in \{0,\cdots,(M-1)\}$. Define
	\begin{IEEEeqnarray}{lCl}
		\label{eq:funcg}	
		g(\mathbf{p},m) &\triangleq& \prod_{i=0}^{N-1} (p_{i})^{m_i} \cdot (\bar{p}_{i})^{1-m_i} \quad ,
	\end{IEEEeqnarray}  	
	where $\mathbf{p}$ is a  vector and $m_i, \forall i$ are the binary coefficients of $m = m_{N-1} \cdot 2^{N-1}+\cdots+ m_0 \cdot 2^0$. 	
	Using (\ref{eq:funcg}), (\ref{eq:Pdfm}) can be written as $Pr\{\mathbf{u}_m \mid H_1\} = g(\mathbf{p_d},m)$ and $Pr\{\mathbf{u}_m \mid H_0\} = g(\mathbf{p_f},m)$.
	\subsection{General Fusion Rule}
	Define a fusion rule $\mathbf{x} \triangleq \left[x_0 \cdots x_{M-1}\right]$, where $x_m$ is the conditional probability that the FC generates $u_{fc}=1$ when  $\mathbf{u}_m$ is received, i.e., $x_m \triangleq Pr\{u_{fc} = 1\mid\mathbf{u}_m \}$.  Then, the probability of detection, $P_D$, and probability of false alarm, $P_F$, of the system associated with $\mathbf{x}$ is \cite{Hoballah1989}:
	\begin{equation}
		\label{eq:PDPF}
		P_D(\mathbf{x}) = \sum_{m=0}^{M-1} x_m \cdot g(\mathbf{p_d},m), \quad
		P_F(\mathbf{x}) = \sum_{m=0}^{M-1} x_m \cdot g(\mathbf{p_f},m).	
	\end{equation}
	Note that when $x_m \in \{0,1\}$, $\{P_D(\mathbf{x}), P_F(\mathbf{x})\}$ are not continuous and take only discrete values. However such fusion rules have an advantage of ease of implementation using boolean switching functions. Alternatively, these rules can be mapped to a single or multiple linear threshold equations \cite{Kohavi2009}.
\section{Problem Formulation}
	\label{sec:prob_form}
	Assuming $\mathbf{p_d}, \mathbf{p_f}$ are known \cite{Chair1986},  we define the generalized decision fusion problem (GDFP) as:
	\begin{IEEEeqnarray}{l?ClCl}
		\label{eq:probDefGDFP}		
		\underset{\mathbf{x}}{\text{maximize}} && R_X(\mathbf{x}) \triangleq C_D \cdot P_D(\mathbf{x}) - C_F \cdot P_F(\mathbf{x})  , \nonumber \\
		\text{subject to} && P_F(\mathbf{x}) \quad \le \alpha ,\nonumber \\
		&& x_m \in \{0,1\},\forall m ,
	\end{IEEEeqnarray}
	where $R_X(\mathbf{x})$ is the cummulative objective function with $C_D$, $C_F$ as coefficients and $\alpha$ is the constraint value on the $P_F$.
		
	As $x_m$ in (\ref{eq:probDefGDFP}) can take 2 possible discrete values $\{0,1\}$, a total number of $2^M (=2^{2^N})$ distinct fusion rules $(\mathbf{x})$ are possible resulting in exponential computational complexity\footnote{Complexity is defined as the number of addition and multiplication floating-point operations (flops).} for finding the optimum fusion rule. When $p_{d_i}>0.5>p_{f_i},\: \forall i$, the optimum fusion rule is \textit{positive unate}, but complexity remains exponential \cite{Viswanathan1997}.

	However, the GDFP as defined in (\ref{eq:probDefGDFP}) is in the form of a classical $0-1$ Knapsack problem (KP) \cite{Kellerer2004} and has a solution using dynamic programming \cite{ErikDemaine2011,Bellman1972} with worst case complexity in polynomial time. The $0-1$ KP is defined as:
	\begin{definition}[$0-1$ Knapsack Problem (KP) \cite{Kellerer2004}]
		Given a set of $M$ items, each	with a weight and value $\{w_m,v_m\}$ respectively for $0 \le m < M$,	choose a subset of items $S$ such that
		\begin{IEEEeqnarray}{l?ClCl}
			\label{eq:genKP}
			\underset{\mathbf{s}}{\text{maximize}} && V(\mathbf{s})  ,\nonumber \\
			\text{subject to} && W(\mathbf{s})  \le W_{lim},
			\quad s_m \in \{0,1\},\forall m,
		\end{IEEEeqnarray}		
	where $\mathbf{s} \triangleq [s_0 \cdots s_{M-1}]$, $s_m$ is the quantity of item $m$ chosen, $V(\mathbf{s})=\sum_{m=0}^{M-1} s_m \cdot v_m$, $W(\mathbf{s})=\sum_{m=0}^{M-1} s_m \cdot w_m$  and $W_{lim}$ is the total weight limit.
	\end{definition}
	\begin{remark}
		The $0-1$ KP has been used in \cite{Hasan2012} for node selection to optimize the performance in an energy constrained setting. To the best of our knowledge, $0-1$ KP is being used for the first time to solve the hard decision fusion rule.
	\end{remark}	
	\begin{theorem}
		The GDFP defined in (\ref{eq:probDefGDFP}) is a $0-1$ KP (\ref{eq:genKP}).
		\begin{IEEEproof}			
			Define individual objective and constrained value respectively of index $m$ as:
			\begin{IEEEeqnarray}{lCl}
				\label{eq:objR}
				R_M(m) &\triangleq& C_D \cdot g(\mathbf{p_d},m) - C_F \cdot g(\mathbf{p_f},m), \nonumber \\
				P_{F_M}(m) &\triangleq& g(\mathbf{p_f},m).
			\end{IEEEeqnarray}
			Let $v_m = R_M(m)$, $w_m = P_{F_M}(m)$, $W_{lim}=\alpha$ and $\mathbf{s} = \mathbf{x}$ as in (\ref{eq:probDefGDFP}), then (\ref{eq:probDefGDFP}) is in the form of (\ref{eq:genKP}).									
		\end{IEEEproof}			
	\end{theorem}
	We now show that various new and existing problems in the literature are special cases of the GDFP. Therefore solution to these problems can be obtained from the solution of GDFP. 	
	\subsection{Special cases of GDFP - New Problems}	
	We define a new count based fusion rule ($\mathbf{y}$) at FC where $u_{fc}$ is decided based on the count of sensors reporting $H_1$. Then, $\mathbf{y} \triangleq [y_0 \cdots y_N]$, where $y_{k} \triangleq Pr\{u_{fc} = 1\mid cnt(m) = k\},0\le k \le N$, and where $cnt(m) \triangleq \sum_{i=0}^{N-1}(m_i)$, henceforth called the \textit{vote count} of $\mathbf{u}_{m}$. For this case:
	\begin{IEEEeqnarray}{lCl}
		\label{eq:relxy}
		\mathbf{x} &=& [y_{cnt(0)} \: y_{cnt(1)} \cdots y_{cnt(M-1)}].
	\end{IEEEeqnarray}
	The individual objective and constraint function respectively for a count $k$ is:
	\begin{equation}
		R_K({k}) \triangleq \smashoperator{\sum_{\forall m :cnt(m)=k}} R_M(m), \quad
		P_{F_K}({k}) \triangleq \smashoperator{\sum_{\forall m :cnt(m)=k}} P_{F_M}(m).
	\end{equation}
	\begin{proposition}[Count based fusion rule (C-GDFP)]
	\label{prop:count}
	The optimum count based fusion rule is a GDFP.
	\begin{IEEEproof}
		Substituting $\mathbf{x} = \mathbf{y}$, $R_X(\mathbf{x}) = R_Y(\mathbf{y})$, $P_F(\mathbf{x}) = P_{F_Y}(\mathbf{y})$, where $R_Y(\mathbf{y}) \triangleq \sum_{k=0}^{N} y_k \cdot R_K({k})$ and $P_{F_Y}(\mathbf{y}) \triangleq \sum_{k=0}^{N} y_k \cdot P_{F_K}({k})$, we get:				
		\begin{IEEEeqnarray}{l?C?l}
			\label{eq:probDefCount}
			\underset{\mathbf{y}}{\text{maximize}} && R_Y(\mathbf{y}), \nonumber \\
			\text{subject to} && P_{F_Y}(\mathbf{y}) \! \le \alpha, \quad  y_k \in \{0,1\},\forall k,
		\end{IEEEeqnarray}
		which is a GDFP.
	\end{IEEEproof}	
	\end{proposition}
	Note $K$-out-of-$N$ rule \cite{Viswanathan1989,Zhang2009} is a special case of this class.
	\begin{proposition}[Discrete Neyman-Pearson (D-NP GDFP)]
	\label{prop:np}
	The optimum fusion rule for maximizing the system $P_D$ with a constraint on $P_F$ is a GDFP.
	\begin{IEEEproof}
		Substituting $C_D = 1, C_F=0$ in (\ref{eq:probDefGDFP}), we get :
		\begin{IEEEeqnarray}{l?ClCl}
			\label{eq:probDefGen}
			\underset{\mathbf{x}}{\text{maximize}} && P_D(\mathbf{x})  ,\nonumber \\
			\text{subject to} && P_F(\mathbf{x}) \! \le \mathbf{\alpha}, \quad
			x_m \in \{0,1\},\forall m .
		\end{IEEEeqnarray}
		which by definition \cite{Trees2001} is a Neyman-Pearson problem.
		\end{IEEEproof}
	\end{proposition}						
	\begin{remark}
		Note that we call this problem as Discrete Neyman-Pearson (D-NP) GDFP. If the constraint on $x_m$ in (\ref{eq:probDefGen}) is relaxed to $x_m \in \mathbb{R}$, $0 \le x_m \le 1$, the problem setup changes to a randomized NP decision fusion problem \cite{Hoballah1986} which has a linear complexity solution.
	\end{remark}
		\subsection{Special cases of GDFP - Existing Problems}	
	\begin{proposition}[Discrete Bayesian CV Problem (D-B GDFP)]
	\label{prop:CVrule}
	The Chair-Varshney (CV) problem for Bayesian criterion in \cite{Chair1986} is a GDFP.
	\begin{IEEEproof}
		Substituting $\alpha = 1$, $C_D = p_1\cdot\left( C_{01}-C_{11}\right)$, $C_F = p_0\cdot\left( C_{10}-C_{00} \right)$ in (\ref{eq:probDefGDFP}), where $C_{ij}$ is the cost of deciding $H_i$ when $H_j$ is true, and $p_j$ is the apriori probability of hypothesis $H_j$, for $i,j \in \{0,1\}$, we have
		\begin{IEEEeqnarray}{l?Cl?Cl}
			\label{eq:probDefGDFP-BC}
			\underset{\mathbf{x}}{\text{maximize}} && C_D \cdot P_D(\mathbf{x}) - C_F \cdot P_F(\mathbf{x}) , \nonumber \\
			\text{subject to} && P_F(\mathbf{x}) \! \le 1 , \quad
			x_m \in \{0,1\},\forall m.
		\end{IEEEeqnarray}
		Since $P_F(\mathbf{x}) \: \le 1$ is not a constraint, (\ref{eq:probDefGDFP-BC}) defines a decision fusion problem for the Bayesian cost function \cite{Trees2001}.		
		Substituting the values $C_D = p_1$ and $C_F=p_0$ in (\ref{eq:probDefGDFP-BC}), completes the proof.
	\end{IEEEproof}
	\end{proposition}
	\begin{proposition}[D-B GDFP with homogenous sensors (HM D-B GDFP)]
	\label{prop:Thomrule}					
		The problem proposed by Thomopoulos et al. \cite{Thomopoulos1987} in CV setting for homogenous sensors with $p_{d_i}=p_d, \:p_{f_i}=p_f,\forall i$ and $p_d > p_f$ is a GDFP.
	\begin{IEEEproof}
		Substituting $C_D = p_1$, $C_F=p_0$, $\mathbf{p_d} = p_d\cdot\mathbf{1}$, $\mathbf{p_f} = p_f\cdot\mathbf{1}$ in D-B GDFP of (\ref{eq:probDefGDFP-BC}) where vector $\mathbf{1} \triangleq \mathbf{1}_{1\text{x}N}$ is a row vector with all elements as $1$, is a special case of CV Problem \cite{Thomopoulos1987}.
	\end{IEEEproof}
	\end{proposition}
	\begin{figure}[!t]
		\centering
		\includegraphics[width=3.25 in]{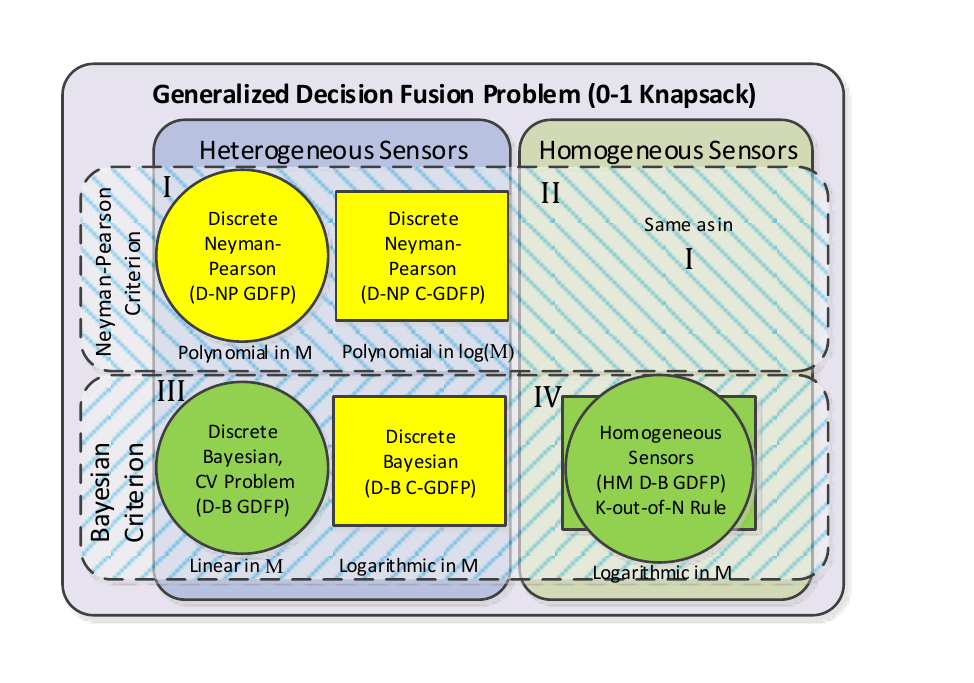}
		\caption{Pictorial representation of GDFP special cases and solution complexity}
		\label{fig_GDFP_mapping}
	\end{figure}
	Figure \ref{fig_GDFP_mapping} pictorially summarizes the GDFP special cases with their settings and proposed solution complexities. Each setting \{I,II,III,IV\}, is a unique combintation of type of sensors \{heterogeneous, homogeneous\}  and test criterion \{NP, Bayesian\}. The distinct shapes \{circle, rectangle\} identify the fusion rule employed \{GDFP, C-GDFP\}, the colours \{yellow, green\} indicate nature of the problem \{new, existing\} respectively. The overlapping of the shapes in quadrant IV indicates that GDFP converges with C-GDFP for this special case.
\section{Solution for GDFP}
	\label{sec:sol_gen}
	Following \cite{ErikDemaine2011,Bellman1972}, we use the dynamic programming concepts to provide a recursive equation and an algorithm that searches for the GDFP optimum fusion rule in polynomial time.
	
	Define a problem $T(a,b)$ as:	
	\begin{equation}
		\label{eq:DP_recur01}
		T(a,b) \triangleq \left\{
		\begin{array}{rl}
		\underset{\mathbf{x^a}}{\text{maximize}} & R_X(\mathbf{x^a}),  \\
		\text{subject to} & P_F(\mathbf{x^a})  \le b,  \\
		& x_m \in \{0,1\},0 <m \le a,
		\end{array}
		\right.
	\end{equation}
	where the integer variable $a$ represents the largest index of the partial fusion vector $\mathbf{x^a} \triangleq [x_0 \cdots x_{a-1}, x_{a}],\: 0<a\le M-1$ and variable $b, \: 0\le b \le \alpha$ is the constraining value. Then problem $T(M-1,\alpha)$ represents the GDFP of (\ref{eq:probDefGDFP}).

	Using (\ref{eq:objR}), we split (\ref{eq:DP_recur01}) recursively as:
	\begin{IEEEeqnarray}{lcl}
		\label{eq:DP_recur02}
		T(a,b) &=& \text{max} \{ R_M(a) + T(a-1,b-P_{F_m}(a)), T(a-1, b) \}, \nonumber \\
		\\
		\text{where,}  \nonumber \\
		T(0,b)	&=&
		\begin{cases}
			0 & \text{for }  0 \le b < P_{F_m}(0),\\
	    	\text{max}\{0,R_M(0)\} & \text{for } P_{F_m}(0) \le b \le \alpha. \label{eq:DP_init}
		\end{cases}	
	\end{IEEEeqnarray}
	Equation (\ref{eq:DP_recur02}) compares and chooses the maximum of the sub-results with and without the contribution from the $a^{th}$ element while satisfing the constraint value.
	
	The key algorithmic approach that reduces the computational complexity is to solve the problem bottom-up by reusing the sub-results.
	To facilitate easy storage and retrieval of the sub-results of (\ref{eq:DP_recur02}), a two dimensional array indexed by values of $a$ and $b$ is used. While $a$ is already an integer variable, the real variable $b$ is mapped onto an integer variable $I_b \triangleq \lfloor C\cdot b+\frac{1}{2} \rfloor$, where $C$ is a sufficiently large scaling factor ($C=10^5 \text{ for GDFP and } 10^3 \text{ for C-GDFP}$ is used) and $0\le I_b \le I_{\alpha}$. Similarly $P_{F_m}(m),\forall m$ is mapped onto the same integer scale as,  $P_{F_m}[m] \triangleq \lfloor C\cdot P_{F_m}(m)+\frac{1}{2}  \rfloor, \forall m$. A two dimensional array $T[M,I_{\alpha}]$ is then used for storage.

	Algorithm \ref{algo01} uses the initial values given in (\ref{eq:DP_init}) and populates the sub-results into the array by looping on integer variable $a$ (line \ref{algo:loopa}) and $I_b$ (line \ref{algo:loopb}). By the end of the iterations, array location $T[M-1,I_{\alpha}]$ is populated with the maximized objective value of (\ref{eq:probDefGDFP}).
	The array is then back tracked to identify and mark the contributing indices $a$ to form the optimum fusion vector $\mathbf{x}^*$. The \textit{Backtrack} method (line \ref{algo:backtrack}) involves memory access operations and does not require any flops.
	
	This algorithm takes a maximum of $3$ flops to compute each sub-result (line \ref{algo:start} to \ref{algo:end}) and hence a total of $3 \cdot I_{\alpha} \cdot M$ flops to compute the solution for GDFP in the worst case which was originally an exponential complex problem ($2^M$).
	\begin{algorithm}
		\caption{Solution to GDFP}\label{algo01}
		\begin{algorithmic}[1]
			\State \text{Initialize } $T[0,0:I_{\alpha}]$ \text{with } (\ref{eq:DP_init})
			\For{$a \gets 1,(M-1)$} \label{algo:loopa}
			\For{$I_b \gets 1,I_{\alpha}$} \label{algo:loopb}
			\If{$P_{F_m}[a] \le I_b$} \label{algo:start}
			\State$	T[a,I_b] = \text{max} \{ T[a-1, I_b],$
			\State$R_M(a) + T[a-1,I_b-P_{F_m}[a]] \:\} $						
			\Else
			\State$T[a,I_b] = T[a-1, I_b]$			
			\EndIf \label{algo:end}
			\EndFor
			\EndFor
			\State $\mathbf{x}^* \gets \text{BackTrack}(T)$ \label{algo:backtrack}
			\State \textbf{Return} $\mathbf{x}^*$
		\end{algorithmic}
	\end{algorithm}		
	We now show that the proposed GDFP polynomial complexity solution can be further simplified using the settings of the  special cases.	
	\subsection{Solution for C-GDFP}
	The solution for this case (classified in Proposition \ref{prop:count}) is same as the main solution. However, as the length of the fusion vector $\mathbf{y}$ is reduced to $N+1$ the computational complexity in the worst case is reduced to $3 \cdot I_{\alpha} \cdot (log(M)+1)$ flops.
	\subsection{Solution for D-B GDFP}
	For this case (classified in Proposition \ref{prop:CVrule} and \ref{prop:Thomrule}), the constraint $P_F(\mathbf{x}) \le 1$ does not effect the GDFP solution space and can be discarded. As a result, the constraint variable $b$ in (\ref{eq:DP_recur02}) is discarded and the initial values given in (\ref{eq:DP_init}) changes to $T(0) = \text{max}\{0,R_M(0)\}$. The loop corresponding to the variable $I_b$ on line \ref{algo:loopb} in Algo.\ref{algo01} is also discarded, and as a result only $1$ flop is required to compute a sub-result. Consequently, a total of $M$ flops are required as in \cite{Chair1986}.
	\subsection{Solution for HM D-B GDFP}
	For the case of D-B GDFP with homogenous sensors (classified in Proposition \ref{prop:Thomrule}), using (\ref{eq:funcg}) we have $g(p\cdot\mathbf{1},m) = (p)^k \cdot (\bar{p})^{N-k}$, where $k = \sum_{i=0}^{N-1} m_i$.	Further substituting this in (\ref{eq:objR}),  we have the individual objective function as:
	\begin{equation}	
	\label{eq:objRh}
	R_M(m) = p_1 \cdot (p_d)^k \cdot (\bar{p_d})^{N-k} - p_0 \cdot (p_f)^k \cdot (\bar{p_f})^{N-k}.
	\end{equation}
	Note that $R_M(m)$ is dependent on the \textit{vote count} $k$.
	As a result, all indices with the same \textit{vote count} $k$ have the same $R_M(\cdot)$ value and consequently the same $x_m$ value in the optimum fusion vector $\mathbf{x}^*$ for D-B GDFP in (\ref{eq:probDefGDFP-BC}).
	In this case, the structure of $\mathbf{x}^*$, is similar to the structure of $\mathbf{x}$ in (\ref{eq:relxy}), implying HM D-B GDFP  is a C-GDFP for Bayesian that is obtained by using $R_K({k}) = \binom{N}{{k}} \cdot R_M(m)$ and $\alpha=1$ in the C-GDFP of (\ref{eq:probDefCount}).
	For this Bayesian case, the loop on variable $a$ (line \ref{algo:loopb} of Algo.\ref{algo01}) has a reduced length of $N+1$. As a result, a total of $log(M)+1$ flops are required.
	\subsection{Solution for HM D-B GDFP with $p_d > p_f$}
	\begin{lemma}
		\label{lemma:voterule}
		 Under the assumption $p_d > p_f$, if $R_K({k}) > 0$, then  $R_K({k}+1) > 0$.
		 \begin{IEEEproof}
		 	Given $R_K({k}) > 0$, $\implies R_M({m}) > 0,$ where $cnt(m)=k$. Under the assumption $p_d > p_f$, we have $\frac{p_d}{p_f} > 1$, $\frac{\bar{p_f}}{\bar{p_d}} > 1$, and using (\ref{eq:objRh}), we have:
			\begin{equation*}
				\dfrac{p_1 \cdot (p_d)^{k} \cdot (\bar{p}_d)^{(N-{k})}}{p_0 \cdot (p_f)^{k} \cdot (\bar{p}_f)^{(N-{k})}} \cdot \frac{p_d}{p_f} \cdot \frac{\bar{p_f}}{\bar{p_d}} > 1,
				\implies R_K({k}+1) > 0. \nonumber
			\end{equation*}
		 \end{IEEEproof}		 	
	\end{lemma}			
	Using Lemma \ref{lemma:voterule}, the HM D-B GDFP can be further simplified in this case as: $\underset{K^*}{\text{maximize}} \sum_{k=K^*}^{N} R_K(k)$ which is a $K$-out-of-$N$ voting rule, where $K^*$ is the smallest integer for which $R_K(K^*)>0$. For $p_0=p_1=\frac{1}{2},\: K^*$ is derived from (\ref{eq:objRh}) as $K^* = \lceil \frac{N}{1+\beta} \rceil$, as in \cite{Hoballah1989,Zhang2009} where $\beta = ln\frac{p_f}{p_d}\:/\:ln\frac{\bar{p_d}}{\bar{p_f}}$ and where, $\lceil\cdot\rceil$ is the ceiling function.
\section{Numerical results and Discussions}
\label{sec:num_res}
	\begin{figure}[!t]\vspace*{-1mm}
		\centering
		\includegraphics[width=3 in]{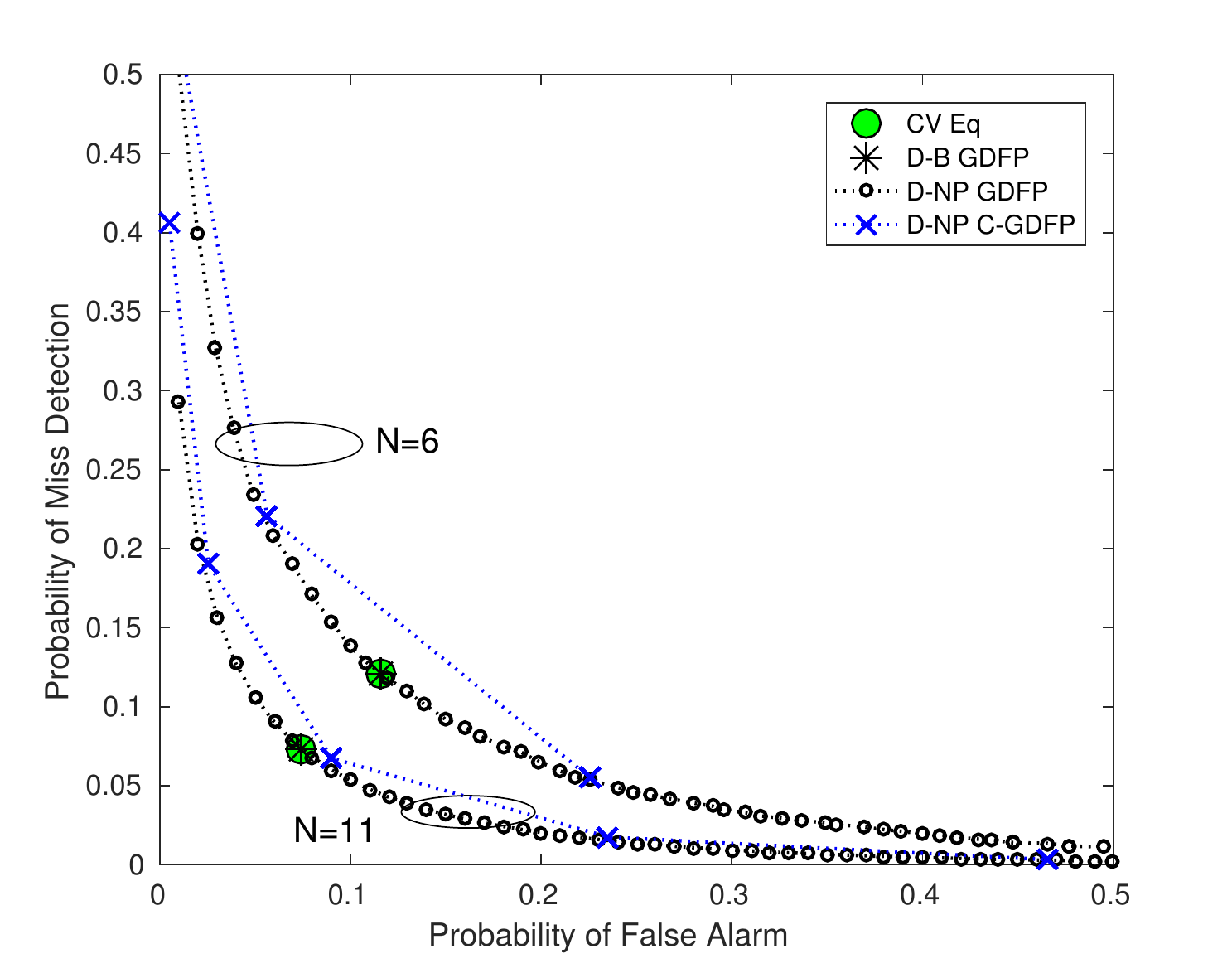}
		\caption{Numerical results for heterogenous sensors}
		\label{fig_sim01}
	\end{figure}
	Figure \ref{fig_sim01} plots the optimum error pairs $P_F^*$ Vs $P_M^* (= 1-P_D^*)$ (labelled D-NP GDFP) obtained using Algo.\ref{algo01} for GDFP under discrete NP test that are computed by varying the allowed limit $\alpha$ from ($0.05\le\alpha\le0.5$) in steps of $0.01$.
	The $p_{f_i}$ and $p_{d_i}$ of each sensor is taken as ($p_{f_i} < 0.5 < p_{d_i}$).
	Note that not all points on this curve are achievable as the $\{P_D,P_F\}$ of the system in (\ref{eq:PDPF}) is discrete. Also note that the plot of optimum error pairs for count based GDFP (labelled D-NP C-GDFP) is piece-wise linear and closely follows the curve D-NP GDFP with fewer achievable points. This implies a few error points with a small performance trade-off are acheivable using C-GDFP in (\ref{eq:probDefCount}) with much lesser computational complexity as listed in Table \ref{tab:flops}. A discrete Bayesian optimum error pair for CV problem assuming $p_0=p_1=\frac{1}{2}$ is computed using linear equation of \cite{Chair1986} (point labelled \textit{CV Eq}) and Algo.\ref{algo01} (point labelled \textit{D-B GDFP}). These coincide with a single error pair achievable by D-NP GDFP.
	\begin{table}
	\renewcommand{\arraystretch}{1.8}
	\caption{Computational complexity of Algo.\ref{algo01} in flops for $M = 2^{11}$}
	\label{tab:flops}
	\centering
		\begin{tabular}{c|c|c|c}
			\hline
			$\alpha$ & Exponential & GDFP & C-GDFP \\
			\hline \hline
			$0.1$  & $\approx 10^{616}$ & $608.4\times 10^5$ & $3.3\times 10^3$ \\
			\hline
		\end{tabular}
	\end{table}
	As, the proposed algorithm requires the constraining values to be linearly mapped onto an integer scale, it acts as a limitation for NP for low precisions. Algorithms based on branch and bound technique \cite{Martello2000} that overcome this limitation are further topics of research.
\vspace*{-1mm}\section{Conclusions}
	\label{sec:con}
	A generalized decision fusion problem is formulated for NP and Bayesian criterion. The proposed GDFP is shown to be in the form of $0-1$ Knapsack problem and results in a polynomial time worst case complexity. A new count based fusion rule has been identified, with significant reduction in complexity and a small penalty on the performance. The GDFP can potentially help uncover more special cases for NP, Bayesian and other criterions, with lower complexity. Conversely, a few special cases of $0-1$ KP have been identified which have significantly lower complexity.
\IEEEtriggeratref{13} \newpage	
\bibliographystyle{IEEEtran}
\bibliography{IEEEabrv,15-07-Generalized_Decision_Fusion}
\end{document}